\documentclass{iopart}
\usepackage{graphicx}

\begin{document}

\title{Space-Based Gravity Detector for a Space Laboratory }
\author{L.V.Verozub}

\address{Kharkov State University,Kharkov 310077, Ukraine}
\begin{abstract}
A space-based superconducting gravitational low-frequency wave detector 
is considered.
Sensitivity of the detector is sufficient to use the detector as a partner
of other contemporary low-frequency detectors like LIGO and LISA. 
This device can also be very useful for experimental study of
other effects predicted by theories of gravitation.
\end{abstract}
\maketitle
\section{Introduction}
The problem of the registration of gravitational waves and investigation
of their properties is too difficult
to suppose that  realization of the
projects like LIGO  will immediately solve all problems.
When the detectors with high sensitivity will begin operate, the problems 
of identification of the
signals will then appear. Under the surcumstances the using of other types
of the detectors would be very useful.
One of them is considered in this paper 
 ( See also \cite{Verozub},
\cite{Ozerov}) 

\section{ A Peculiarity of Magnetically Interacting Superconducting
Solenoids}

The detector is based on the 
a peculiarity of  magnetically interacting superconducting solenoids
that was discovered theoretically by a group of Ukrainian engineers 
\cite{Kozorez}. 

Consider a pair of the superconducting solenoids $A$ and $B$ in line
(Fig. 1)
in the weightless state . If the
solenoids carry the persistent currents $I_{1}$ and $I_{2}$ ,
it can be shown that the complete energy of the system is given by 
\begin{equation}
U=(L_{2}Q_{1}^{2}-2MQ_{1}Q_{2}-L_{1}Q_{2}^{2})/2D.  \label{U}
\end{equation}
where $L_{1}$ and $L_{2}$ are the solenoids inductances , $M$ is the 
mutual
inductance of the solenoids that is the function of the distance
$x$ between the solenoids centers, $D=L_{1}L_{2}-M^{2},$ $Q_{1}$and $Q_{2}$ 
are the constant fluxes in the solenoids.

The solenoids are attracted by the Ampere force $F=-\partial W/\partial x
$ affecting the solenoids. 

The peculiarity of this interaction is that  at the condition
 $Q_{1} << Q_{2}$ the Ampere force change 
its sign at some distance $x=x_{0}$ between the solenoids, 
and the function $W(x)$ has the minimum  . At this position 
the solenoids in the weightless state are in a week stable equilibrium 
condition.

A typical dependence of the force $F$ of the distance $x$  
(in CGS units) is given in 
Fig. 2 . The parameters of the solenoids are : the inductances are $1.15Hn.$
, the lengths are $5cm.$, the radiuses are $5cm.$, $Q_{1}=1.15Wb$ ,
$Q_{2}=Q_{1}/100$ .

\begin{figure}[htb]
\centering
\includegraphics[width=45mm,height=30mm]{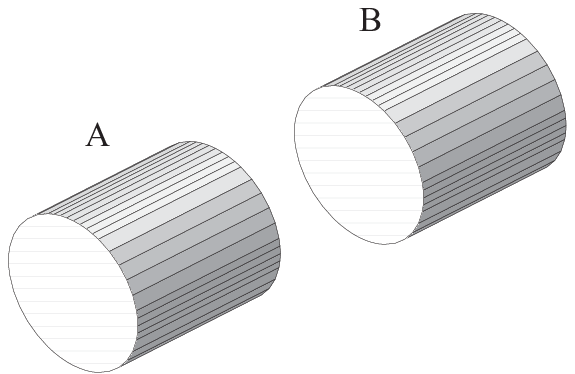}
\hfill
\includegraphics[width=55mm,height=35mm]{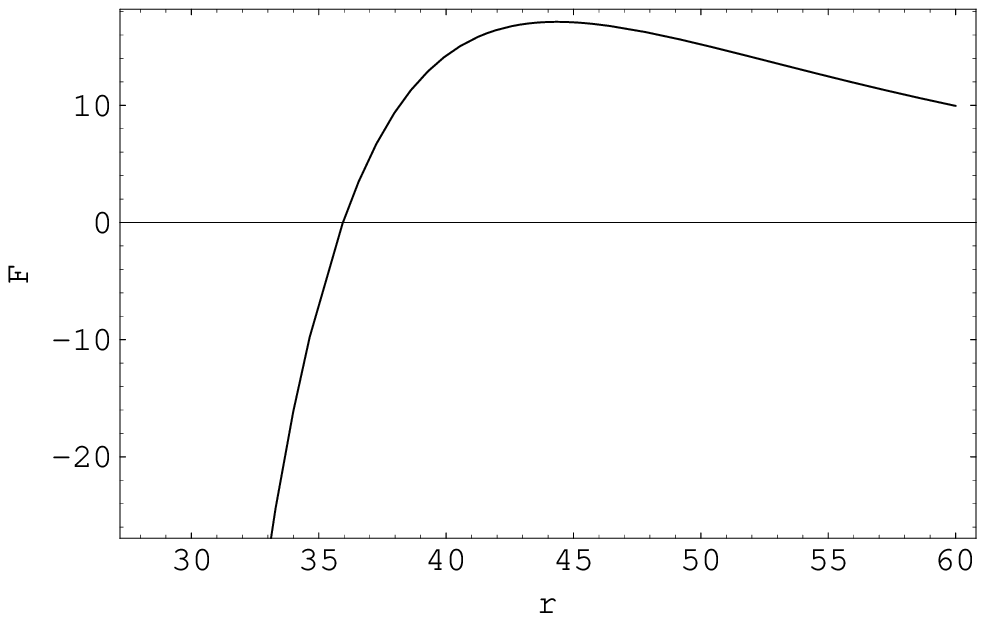}
\\
\parbox[t]{55mm}{\caption{The solenoids in weightless state}} 
\hfill
\parbox[t]{55mm}{\caption{The Ampere force between the 
solenoids }}
\label{solenoids}
\end{figure}

Our experimental investigations  confirm this result \cite{Ozerov}.

\section{ The Magnetically Coupled Solenoids as a Detector of Tidal
Accelerations}

Consider the properties of the system, formed by a pair of superconducting
solenoids in weightless state, in the field of a gravitational wave with the
frequency $\nu $ extending during the time interval $t_{0}$ perpendicularly
to $AB$ - direction. This system is a nonlinear oscillator, the small
oscillations of which in $AB$ - direction are described by the differential
equation 
\begin{equation}
m\ddot{q}+R(\dot{q})+pq=f(t).  \label{eqoscillator}
\end{equation}
In eq.(\ref{eqoscillator}) $q=x-x_{0}$ is a small deviation from the
equilibrium position , $\dot{q}=\partial q/\partial t$ , $\ddot{q}=\partial 
\dot{q}/\partial t$ , $R(\dot{q})$ is the air resistance, $p=U^{\prime
\prime }(x_{0})$ is the stiffness of the ''magnetic spring'', $%
f(t)=ma_{g}\sin (\omega t)$ at $0<t<t_{0}$ and $f(t)=0$ at $t>t_{0}$ , $m$
is the mass of the system , $\omega =2\pi \nu $ and $a_{g}=\omega
^{2}hx_{0}/2$ is the amplitude of the tidal acceleration, caused by the
gravitational wave.

If 1-type superconductors are used in the solenoids, the air resistance $R(
\dot{q})$ is the key cause of the oscillations damping in the given system.
.For an ideal gas the function $R(\dot{q})=-b\dot{q}|\dot{q}|$, where $b$ ,
is a constant .

The typical sizes are: $R=5-10cm$ ,$x=30-40cm$ , The proper frequency is
less than $\ 1Hz$ and can reach $10^{-4}Hz.$

Fig. 4  shows the response $q(t)$ (in CGS units system) of the 
detector to the gravitational
wave with the dimenssionless amplitude $h=10^{-20}$ that shows Fig. 3.  

\begin{figure}[htb]
\centering
\includegraphics[width=55mm,height=35mm]{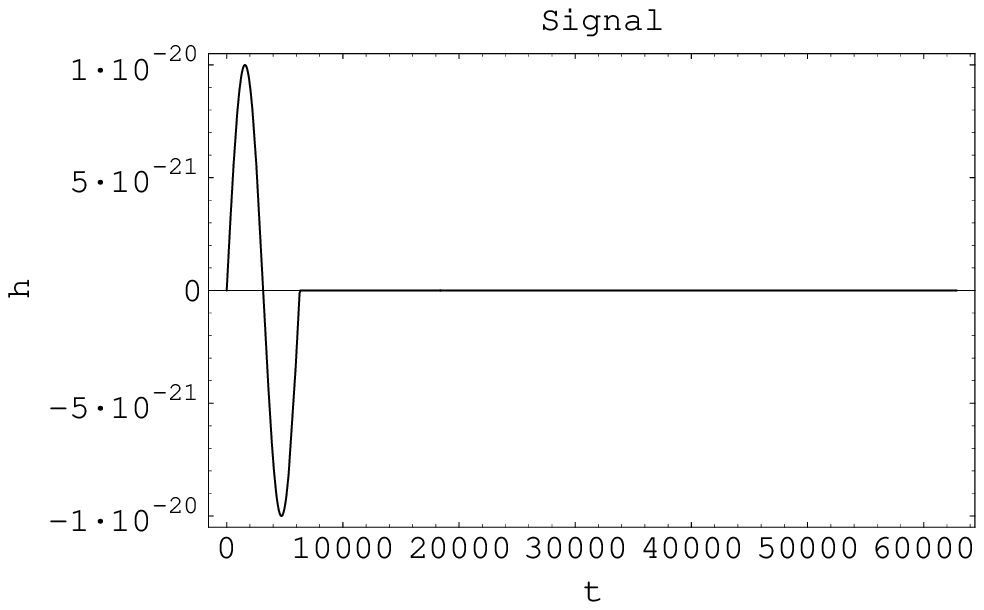}
\hfill
\includegraphics[width=55mm,height=35mm]{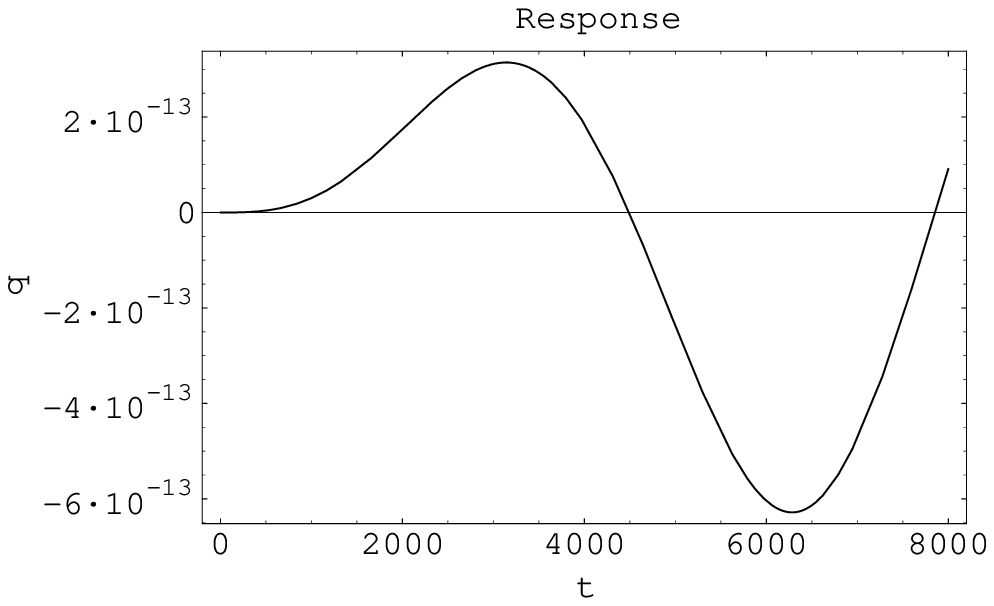}
\\
\parbox[t]{55mm}{\caption{The signal}} 
\hfill
\parbox[t]{55mm}{\caption{The detector response  $q(t)$}} 
\label{eps1}
\end{figure}

\section{Noises}

The detector under study is a nonlinear oscillator. 
Using the Chandasekhar results concerning a linear
oscillator and the method of the statistic linearization for the function 
$R(\dot{q})$ we have found that the the root-mean-square magnitude of the thermal
fluctuations is : $<q^{2}>^{1/2}=\sigma _{q}$ where 
\begin{equation}
\sigma _{q}^{2}(t)=\frac{4b^{2}}{(2\pi )^{1/2}m\omega _{0}^{2}}\frac{kT}{m}%
t^{2}\biggl [1-\frac{\sin ^{2}(2\omega _{0}t)}{2\omega _{0}^{2}t)^{2}}\biggr
]
\end{equation}

For example, if $m=10^{4}gm$ , $\nu _{0}=0.01s$, $t=100s$ , $b=10^{-4}gm/cm$%
, then the root-mean-square magnitude of the thermal fluctuations is : $%
<q^{2}>^{1/2}=10^{-20}cm$. Thus, in spite of the small mass the detector has
a very low level of the thermal noise.

The inhomogeneity of the Earth and spacecraft gravitational fields leads to
more serious problems. The tidal acceleration of the solenoids due to
inhomogeneity of the Earh gravity field is much larger than the useful
signal . This problem can be solved only by choosing a geostationary or a
very distant from the Earth of the spacecraft orbit.However, it should be 
noted
that just this fact
 allows to use the detector as an excellent gravity gradiometer
for the investigation of the Earth gravity field . 

It is necessary to take into account the fluctuations in the gravity
gradient within the spacecraft , instability of the temperature of the
solenoids, posibilty of a "piezoeffect" in superconductors 
\cite{Verozub84}, \cite{Verozub88}, \cite{Peng1}-\cite{Peng3}
and other measuremens noises. 
As a result, we estimate the
minimal relative shift of the solenoids available for the measurements as
the magnitude of the order of $10^{-18}cm$ .

We mean that the solenoids $A$ and $B$ are inside a supeconducting shield.

Consider briefly the physical principles of the solenoids relative shift
measurement caused by the gravitational-wave bursts. The idea is to measure
the change in the proper magnetic flux of one of the solenoids by the SQUID
attached to the solenoids.
A superconducting
quantum- interferometer device (SQUID)  is attached to the solenoid 
and coupled to the other one inductively by a flux transformer .
The results of the SQUID measurements are transmitted by radio by means of a
conversion ''voltage - frequency'' and by using an isotropic active antenna.
Such a method of the solenoids shift measurement is insensitive to
micrometeorite impacts and other forces affecting the spacecraft.

\section{Conclusion}
The detector under consideration is a very promising device for gravitational
waves and other gravitational space-based experimental projects. 
(See also the paper \cite{Karim}).
It is a 
technically  difficult project that needs further detail study.


\section{References}
\begin{thebibliography}{10}
\bibitem{Verozub} L.V. Vrozub (1996) {\it GRG}, {\bf 28}, 77. 
\bibitem{Ozerov}  V.T. Petrenko, etc (1977) {\it IEEE Transection on Appl. 
Superconductivity}  {\bf 7}, 1933.
\bibitem{Kozorez} V.V. Kozorez (1981) {\it Dynamic systems of magnetically
interacting solids}, Kiev (In Russian). 
\bibitem{Verozub84} L.V. Verozub (1984) {\it Sov. Journ. Low Temp. Phys},
{\bf 10}, 103.
\bibitem{Verozub88} L.V. Verozub (1988) {\it Sov. Journ. Low Temp. Phys},
{\bf 14}, 289.
\bibitem{Peng1} H.Peng (1990) {\it GRG}, {\bf 22}, 33.
\bibitem{Peng2} H.Peng and B.Peng (1990) {\it GRG} {\bf 22}, 45.
\bibitem{Peng3} H.Peng and G.Torr (1990) {\it GRG}  {\bf 22}, 53.
\bibitem{Karim} M.Karim (1980) {\it J.Phys. A: Math. Gen.} {\bf 13}, 213. 
\end{thebibliography}
\end{document}